# Extremely efficient permutation and bootstrap hypothesis tests using R

Christina Chatzipantsiou[1,2], Marios Dimitriadis[2], Manos Papadakis[2] and Michail Tsagris[2]

[1] School of Medicine, [2] Department of Computer Science,
University of Crete

Abstract

Re-sampling based statistical tests are known to be computationally heavy, but reliable when small sample sizes are available. Despite their nice theoretical properties not much effort has been put to make them efficient. In this paper we treat the case of Pearson correlation coefficient and two independent samples t-test. We propose a highly computationally efficient method for calculating permutation based p-values in these two cases. The method is general and can be applied or be adopted to other similar two sample mean or two mean vectors cases.

**Keywords:** Computational efficiency, permutation, bootstrap, hypothesis testing.

## Introduction

Computer intensive techniques, such as MCMC , bootstrap and permutation, are known to require computational power. There has been too much of research effort on trying to make MCMC more efficient, either by means of programming or statistical theory. Bootstrap and permutation are on the other hand are not that heavy. Both of them are very useful when performing simulation studies or even analyzing real datasets with small sample sizes and strange shapes of distributions (mixtures for example), or with statistics whose distribution is not known or is too complex to work with.

Two examples where the necessity for, efficient, computational statistics is apparent, are Bayesian networks (Neapolitan, 2003) and feature selection algorithms (Ttsamardinos, 2003). When applied to biological data, which usually have a small number of observations, re-sampling techniques is an advisable strategy. Pearson correlation is a cornerstone in Bayesian network learning (Neapolitan,

2003). Welch's t-test (Welch, 1938) on the other hand is very frequently utilized in gene expression data to discover which genes are differentially expressed, or to compare between (unmatched) case control samples. Other cases include multivariate statistics, which involve calculation of statistical functions whose distribution is not known. One such example is distance correlation (Szekely et al., 2007). Hypothesis testing procedures for multivariate data require large sample sizes in order to have a valid behavior of the type I error. (Tsagris, 2017).

With a focus on univariate statistics, Neto (2015) proposed a vectorized function in R (R, 2017), as an alternative to the "for" function, to speed up the calculation of the bootstrap p-value of the hypothesis test of zero correlation. Comparing his permutation approach with other bootstrap functions available in other R packages, Neto reported, in his work, dramatic reduction in execution runtime. Neto (2015) extended his bootstrap version of the two sample Welch's t-test among others.

Moving along the same lines, those of computational savings, we propose a method for a fast bootstrap or permutation based p-value calculation. Our method is extremely efficient even for a large number of permutations or bootstrap samples (high CPU to data-size ratio tasks). Furthermore, it outperforms Neto's vectorization method, showcasing a notable reduction in execution time even by an order of magnitude in some cases (Figure 3). The focus of this paper is on Pearson correlation and two independent samples t-test. However, due to the method being simple, we show another case where it can be extended, James multivariate version of the t-test (James, 1954), which we briefly discuss.

## Bootstrap or permutation calibration

Our focus is on the correlation and the univariate t-test. The James test is also mentioned as a case example, where our method can be applied.

### Pearson correlation coefficient

Given a sample n of paired observations $(x_i, y_i)$, $i = 1, \ldots, n$, the sample Pearson correlation coefficient is calculated as

$$r = \frac{n\sum_{i=1}^{n} x_i y_i - \sum_{i=1}^{n} x_i \sum_{i=1}^{n} y_i}{\sqrt{n\sum_{i=1}^{n} x_i^2 - (\sum_{i=1}^{n} x_i)^2} \sqrt{n\sum_{i=1}^{n} y_i^2 - (\sum_{i=1}^{n} y_i)^2}}. \qquad (1)$$

In order to test whether the true correlation, $\rho$, is equal to 0, the relevant test statistic is given by

$$Z = 0.5 \log\left(\frac{1+r}{1-r}\right)\sqrt{n-3}. \qquad (2)$$

If $\rho = 0$, $Z \sim N(0,1)$ asymptotically, and for small $n$, $N(0,1)$ can be substituted by $t_{n-3}$.

**Welch's t-test**

The two independent samples t-test is used to check whether the means of two populations, out of which the samples have been drawn are equal (null hypothesis). The main and rather strong assumption is that the variances are the same. Welch (1938) proposed an alternative to the t-test for the case of unequal variances, whose test statistic is given by

$$T = \frac{\bar{x}_1 - \bar{x}_2}{\sqrt{s_1^2/n_1 + s_2^2/n_2}}, \qquad (3)$$

where $\bar{x}_1$ and $\bar{x}_2$ are the two sample means, $s_1^2$ and $s_2^2$ are the two sample variances and $n_1$ and $n_2$ are the two sample sizes. The test statistic (3) is compared against a $t$ distribution with some properly estimated degrees of freedom (Satterthwaite, 1946, Welch, 1947).

**James multivariate t-test**

As an extra case, we also cover the case of the multivariate two independent samples t-test, namely the James test (James, 1954), which is the analogue of Welch's t-test. Similarly to Welch's t-test, James test makes no assumptions about the covariance matrices of the two populations from which the samples were sampled. The test statistic for two d-dimensional samples is

$$T_u^2 = (\bar{X}_1 - \bar{X}_2)^T \left(\frac{S_1^2}{n_1} + \frac{S_2^2}{n_2}\right)^{-1} (\bar{X}_1 - \bar{X}_2), \qquad (4)$$

where $\bar{X}_1$ and $\bar{X}_2$ are the two sample mean vectors, $S_1^2$ and $S_2^2$ are the two sample covariance matrices and $n_1$ and $n_2$ are the two sample sizes. The test statistic (4) is compared against a corrected $X^2$ distribution (James, 1954).

**Permutation and bootstrap based p-values**

As we can see in (1), Pearson correlation coefficient is a function of two vectors of paired observations. The usual permutation based p-value reorders the values of x or of y, thus changing the pairs (Legendre, 2012, Berry, 2016). For every reordering of the values, (2) is calculated. This method is implemented in R by using a "for" loop. Vectorization is a faster alternative, if the number of permutations is the most used in practice, 999, otherwise if this number is 9999 a "for" loop might be a better option, for R, because of larger memory requirements and the need to handle matrices efficiently.

In the context of Bayesian network learning (Neapolitan, 2003), Tsamardinos et al. (2010) suggested a different approach with high computational savings with the $G^2$ test of independence for categorical data. Instead of permuting the data B (e.g. 999 times) times, they calculated the test statistic using 100 permutations. Their average is used as an estimator of the degrees of freedom of the $x^2$ distribution. They also suggest to perform half of the permutations and estimate the probability of having a p-value less than the given significance level. Similarly, it is possible to stop the permutations once the proportion of times the permuted test statistic value exceeds the significance level.

**Neto's (2015) vectorized bootstrap p-value**

Neto (2015) proposed a computationally less heavy calculation of the p-value when using non-parametric bootstrap. We show the Pearson correlation case, and note that similar calculations take place in the case of the Welch's t-test.

Let $n_i^*$ represent the number of times observation $x_{i\_}i$ is selected in the bootstrap sample $X^*$ and $w_i^* = n_i^*/N$, where N is the sample size. Then, the category counts, $\boldsymbol{n}^* = (n_1^*, \ldots, n_N^*)$ of the bootstrap sample $\boldsymbol{x}^*$ are distributed according to the multinomial distribution

$$\boldsymbol{n}^* = N\boldsymbol{w}^* \sim Multinomial(N, N^{-1}\boldsymbol{J}_N), \quad (5)$$

where $\boldsymbol{J}_N = (1, \ldots, 1)^T$ is the unit vector of size N.

As an example, Neto (2015) described his algorithm the case of the sample mean $\hat{\theta} = \frac{\sum_{i=1}^{N} x_i}{N}$.

1. Draw B bootstrap count vectors $\boldsymbol{n}^*$ from (5) using the command "rmultinom" in R.
2. Divide $\boldsymbol{n}^*$ by N to obtain the $N \times B$ bootstrap weights matrix $\boldsymbol{W}^*$.
3. Generate the vector of bootstrap means as $\hat{\theta}_{boot} = \boldsymbol{x}^T \boldsymbol{W}^*$.

For the case of the Pearson correlation coefficient, the vector with the bootstrap correlations is given by

$$r = \frac{(x \cdot y)^T \boldsymbol{W}^* - (x^T \boldsymbol{W}^*) \cdot (y^T \boldsymbol{W}^*)}{\sqrt{(x^2)^T \boldsymbol{W}^* - (x^T \boldsymbol{W}^*)^2} \cdot \sqrt{(y^2)^T \boldsymbol{W}^* - (y^T \boldsymbol{W}^*)^2}}$$

where $x \cdot y = (x_1 y_1, \ldots, x_n y_n)^T$ is the element-wise multiplication of two vectors.

**Efficient calculation of the permutation and bootstrap p-value**

Our efficient permutation and bootstrap based p-value, unlike the former strategies, performs nearly all B permutations. It is currently developed for the Pearson correlation and Welch's t-test. For the Pearson correlation (permutation case) it reorders the values of each vector $[\sqrt{B}]$ times, where $[.]$ denotes the rounding operation. The same idea is used for Welch's t-test (bootstrap case), but with the difference, that for each vector $[\sqrt{B}]$ bootstrap samples are taken. The process is similar for the James test. We create $[\sqrt{B}]$ samples from each sample and store their mean vectors and covariance matrices. The test statistic is then calculated for all $[\sqrt{B}]^2 \simeq B$ combinations of the permuted or bootstrap samples. This simple methodology results in substantial computational gains and the extra benefit is that similar ideas can be used in other settings, for example in calculating the permutation p-value of the distance correlation (Szekely, 2007).

## Time comparisons and statistical validation

For the Pearson correlation and Welch's t-test scenarios we used sample sizes ranging from 10 up to 300 at an increasing step of 10 and 5 different numbers of permutations or bootstrap samples B = (999, 4999, 9999, 14999, 19999). For each combination of them (180 in total), we sampled random vectors 10 times and for each of these times we estimated the time of each function using the package "microbenchmark" (2015) with 100 repetitions.We downloaded Neto's vectorised bootstrap R code https://github.com/echaibub/VectorizedNonParametricBootstrap. For the James test, the sample sizes ranged from 10 up to 100 and the bootstrap samples were B = (999, 4999).

**Vectorized bootstrap versus efficient permutation of Pearson correlation coefficient**

We compared the computational times of our proposed method versus using two functions with a "for" loop in R and using the command "replicate". Figure 1 illustrates the comparison between the aforementioned methods, and the relative performance improvements in terms of runtime.

As a second experimentation we also tested the speed-up factor of the asymptotic p-value versus the permutation based p-value. Figure 2(a) shows that when B=999 and B=4,999 the permutation p-value is less than $\sqrt{B}$ times slower even for large sample sizes. The same is true for larger values of B but for sample sizes less than 150 or 200.

Figure 2(b) shows the number of times Neto's bootstrap implementation (Neto 2015) is slower than our proposed permutation method. We furthermore tested the execution speed differences between our R and C++ implementations. Figure 2(c) shows that for small sample sizes, C++ is 2-3 times faster, but as the sample size grows, the computational cost becomes the same.

As an addition to the simulation study described above, we also performed a comparison of the Welch's t-test on a real gene-expression dataset containing 40 rows (observations) and 54,4675 columns (probesets). The dataset has the GSE15913 accession number in the GEO platform and can be downloaded from BioDataome (Lakiotaki et al.,2018). This dataset was chosen because it contains few samples. Bootstrap is mainly designed for small sample sized data for ensuring the correctness of tests, for example that the nominal selected significance level is close to the actual size of the test. We estimated the time required to perform bootstrap Welch's t-test for all columns. The average over

10 repetitions, using a "for" loop to traverse the whole matrix, four our method was 6 seconds, whereas for Neto's permutation method was 116 seconds (19 times slower).

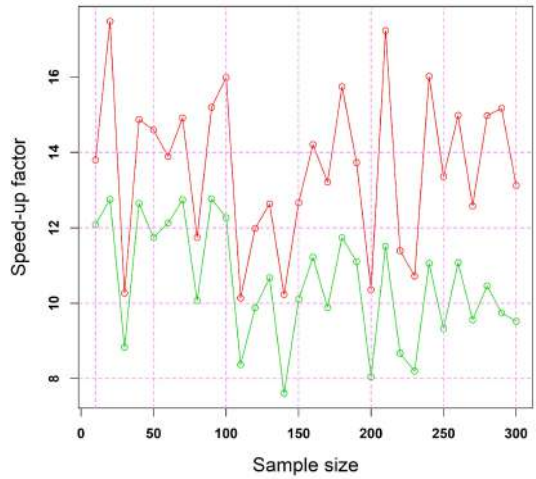

(a) B = 999

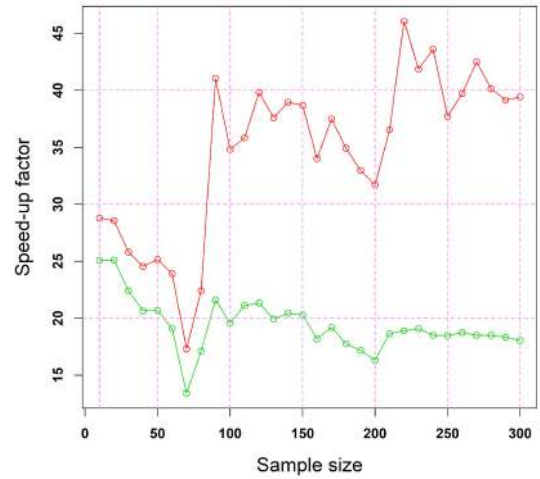

(c) B = 4,999

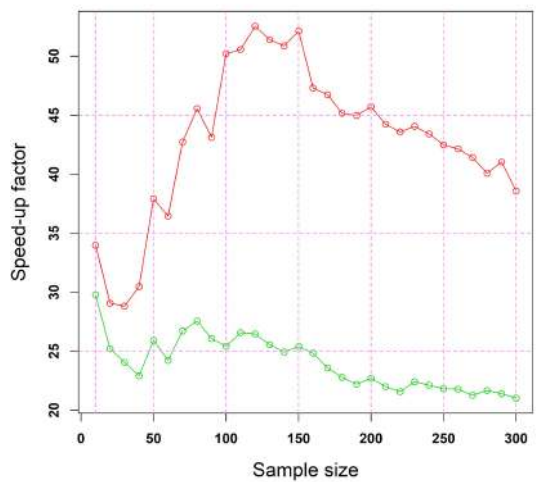

(b) B = 9,999

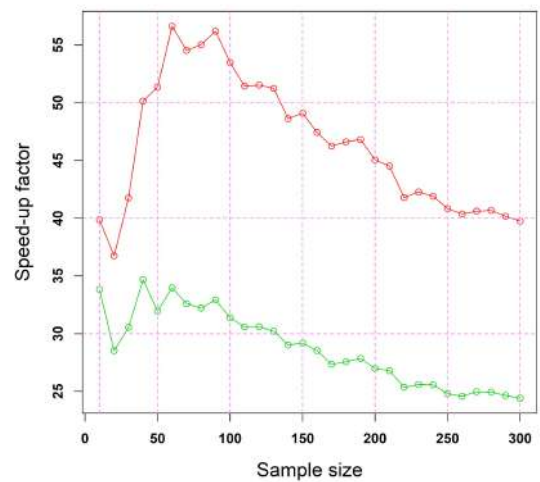

(d) B = 19,999

Figure 1. **Correlation coefficient**. Speed-up factors of some standard ways versions against the new version. Higher numbers indicate higher computational cost. The green line is with "replicate" while the red line is with "for".

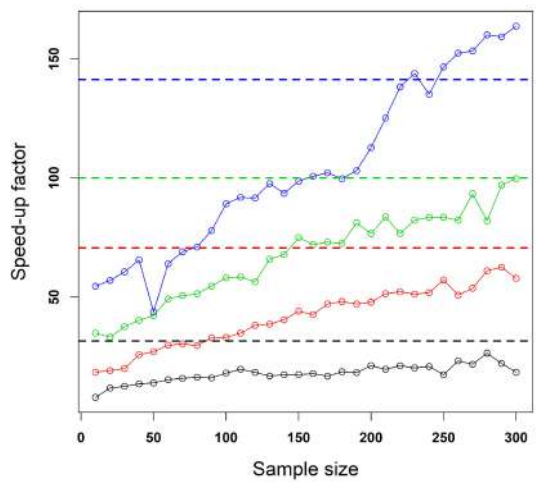 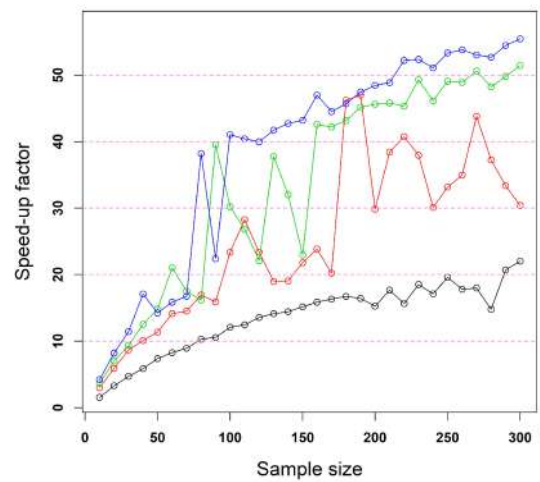

(a) (b)

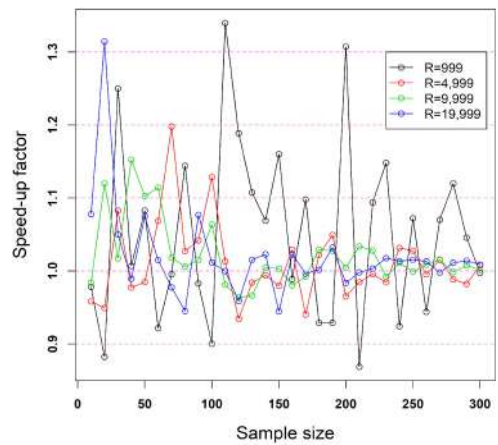

(c)

Figure 2. **Correlation coefficient**. (a) Speed-up factors of the asymptotic p-value versus the permutation p-value. This is an estimate of how much slower the permutation p-value is, relative to the asymptotic p-value. (b) Speed up factor of the vectorized bootstrap (Neto, 2015) versus our permutation implementation. Numbers greater than 1 indicate that our version is faster. (c) Speed up factor of the C++ implementation versus the R implementation. This is an estimate of how slower R is in comparison to C++ for this method.

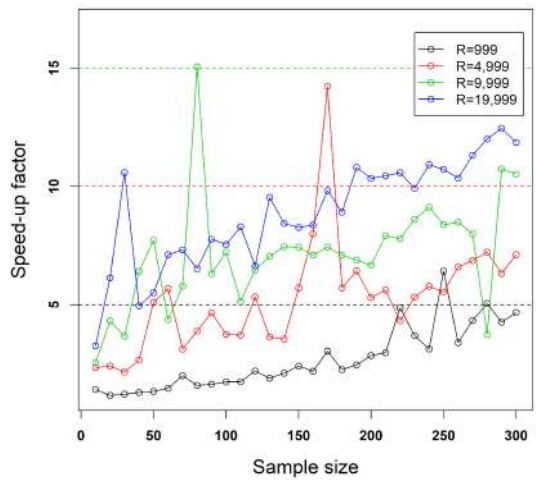 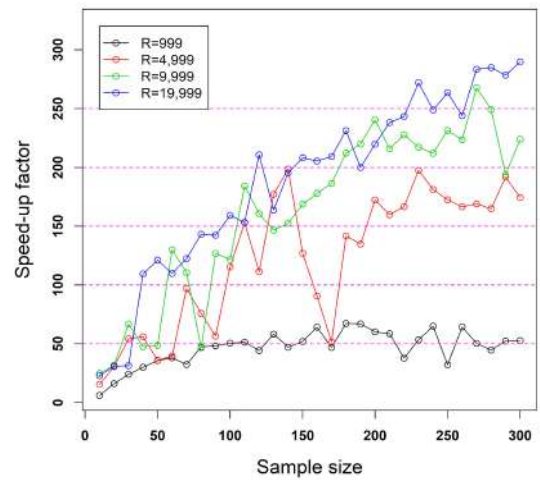

(a)                                                    (b)

Figure 3. **Welch's test**. (a) Speed-up factors of the asymptotic p-value versus the permutation p-value. This is an estimate of how much slower the permutation p-value is, relative to the asymptotic p-value. The built-in command "t.test" was used. (b) Speed up factor of the vectorized bootstrap (Neto, 2015) versus our bootstrap implementation. This is an estimate of how much slower Neto's permutation method is, in comparison to our permutation method.

**Ordinary bootstrap versus efficient bootstrap for James test**

As a third case of application, we examine the James multivariate version of the Welch's t-test. The rationale behind the application of our method for this test is similar to the ones previously described in this work. Figure 4 presents the speed-up factors between the ordinary bootstrap procedure and our proposed efficient bootstrap procedure.

The R code for both cases appears in the Appendix. One can see, the number of "for" loops is more than the usual bootstrap implementation. The difference though lies in the number of calculations. In the ordinary bootstrap implementation, 2B covariance matrices are calculated, B for each sample. Using the efficient method, only $2[\sqrt{B}]$ covariance matrices are calculated. The computational cost does not reduce by $[\sqrt{B}]$ simply because, we use more "for" loops and there are other heavy operations as well, which we cannot avoid, such as inversion of covariance matrices. The number of inversions is $\left[\sqrt{B}\right]^2 > B$, and hence we also spend some time there as well.

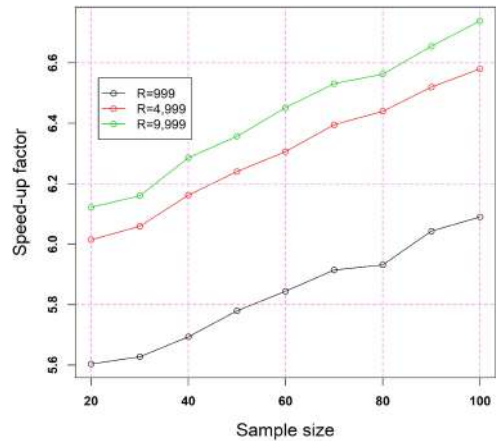

Figure 4. **James test**. Speed-up factors of the efficient versus the ordinary permutation bootstrap for the James test. This is an estimate of how much slower the ordinary bootstrap is, relative to our efficient bootstrap.

**Statistical validation in terms of type I and type II errors**

An equally important aspect we examined was the statistical side. For both the correlation and Welch's t-test we generated data under the $H_0$ hypotheses, of no correlation and equal means respectively. Figure 5 clearly shows that the quantiles of the true distribution are accurately estimated by our proposed method for the correlation and the Welch's t-test. Estimations for the James test at this point were not calculated; Tsagris et al.(2017) compared many hypothesis testing procedures with and without bootstrap calibration and have highlighted the importance of the bootstrap calibration in the multivariate case.

## Conclusions

We presented an efficient methodology to obtain permutation based p-values for the Pearson correlation coefficient, the two sample t-test and extended it the two sample multivariate t-test (James test). Another case where this method can be applied is for the distance correlation, where the p-value is estimated via permutations (Szekely, 2007). In (Bayesian) network construction, when partial correlation and partial distance correlations are employed, this method could help speed-up the process significantly. We have implemented and included our proposed efficient methodology in the R package "Rfast" (Papadakis et al., 2018), as the functions `permcor` and `boot.ttest2`, for the Pearson correlation and Welch ttest respectively. The relative R code is also included in the Appendix.

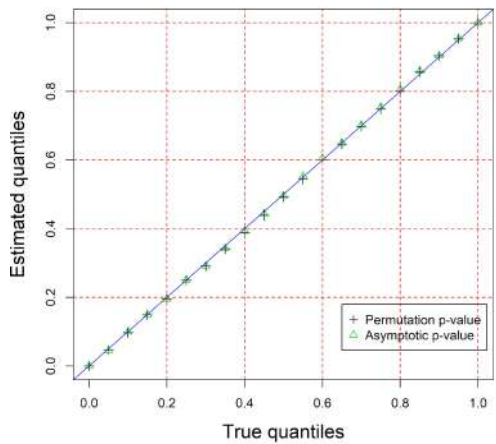

(a)

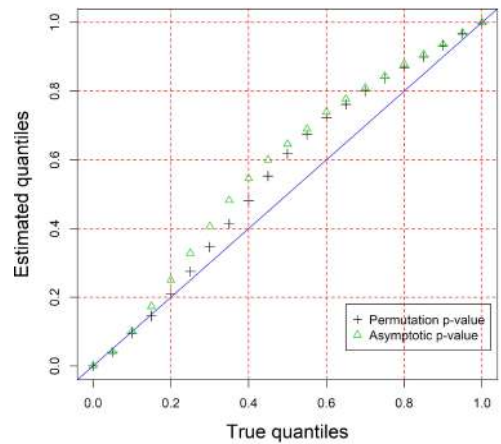

(c)

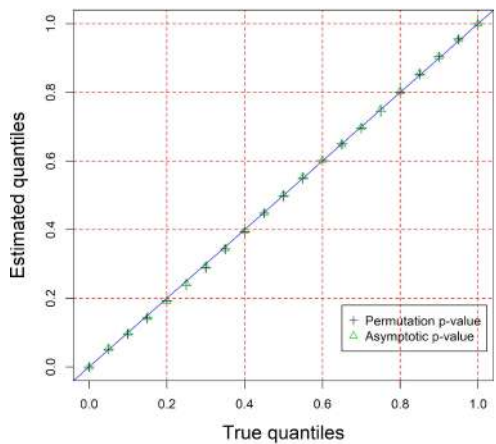

(b)

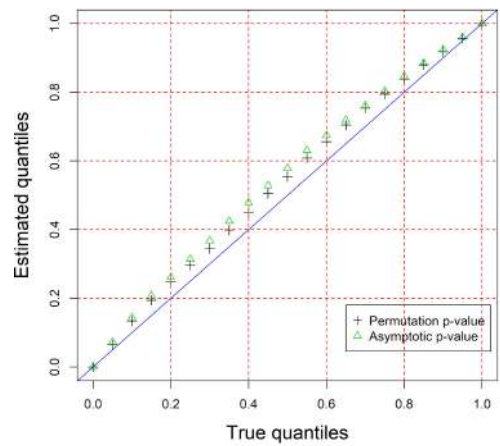

(d)

Figure 5. Estimated quantiles of the distribution of the test statistic under the null hypothesis. (a) + (b) refer to the correlation coefficient, while (c) + (d) refer to the Welch's t-test.

# Appendix

library(Rfast)  ## necessary package to load

### Permutations for correlation coefficient

```
permcor <- function(x, y, R = 999) {
  n <- length(x)    ;   m1 <- sum(x)       ;   m12 <- sum(x^2)
  m2 <- sum(y)    ;   m22 <- sum(y^2)    ;   up <-  m1 * m2 / n
  down <- sqrt( (m12 - m1^2 / n) * (m22 - m2^2 / n) )
  r <- ( sum(x * y) - up) / down
  test <- log( (1 + r) / (1 - r) )  ## the test statistic
  B <- round( sqrt(R) )    ;   xp <- matrix(0, n, B)    ;   yp <- matrix(0, n, B)
  for (i in 1:B) {
     xp[, i] <- sample(x, n)    ;    yp[, i] <- sample(y, n)
  }
  sxy <- crossprod(xp, yp)    ;   rb <- (sxy - up) / down
  tb <- log( (1 + rb) / (1 - rb) )
  pvalue <- ( sum( abs(tb) > abs(test) ) + 1 ) / (B^2 + 1)  ## p-value
  res <- c( r, pvalue )
  names(res) <- c('correlation', 'p-value')
  res
}
```

### Ordinary bootstrap for James test

```r
james <- function(y1, y2, R = 999) {
  p <- dim(y1)[2]    ;   n1 <- dim(y1)[1]    ;   n2 <- dim(y2)[1]
  ybar1 <- Rfast::colmeans(y1)    ;    ybar2 <- Rfast::colmeans(y2)
  dbar <- ybar2 - ybar1   ;    A1 <- Rfast::cova(y1)/n1    ;   A2 <- Rfast::cova(y2)/n2
  test <- as.vector( dbar %*% solve(A1 + A2, dbar ) )
  a1inv <- Rfast::spdinv(A1)        ;        a2inv <- Rfast::spdinv(A2)
  mc <- solve( a1inv + a2inv, a1inv %*% ybar1 + a2inv %*% ybar2 )
  mc1 <-  - ybar1 + mc            ;      mc2 <-  - ybar2 + mc
  x1 <- Rfast::eachrow(y1, mc1, oper = "+")
  x2 <- Rfast::eachrow(y2, mc2, oper = "+" )
  tb <- numeric(R)
  for (i in 1:R) {
    b1 <- sample(1:n1, n1, replace = TRUE)    ;   b2 <- sample(1:n2, n2, replace = TRUE)
    xb1 <- x1[b1, ]    ;     xb2 <- x2[b2, ]
    db <- Rfast::colmeans(xb1) - Rfast::colmeans(xb2)
    Vb <- Rfast::cova(xb1) / n1 + Rfast::cova(xb2) / n2
    tb[i] <- sum( db %*% solve(Vb, db ) )
  }
  ( sum(tb > test) + 1 ) / (R + 1)
}
```

## Efficient bootstrap for James test

```
boot.james <- function(y1, y2, R = 999) {
  p <- dim(y1)[2]      ;    n1 <- dim(y1)[1]     ;    n2 <- dim(y2)[1]
  ybar2 <- Rfast::colmeans(y2)     ;    ybar1 <- Rfast::colmeans(y1)
  dbar <- ybar2 - ybar1   ;   A1 <- Rfast::cova(y1)/n1    ;    A2 <- Rfast::cova(y2)/n2
  test <- as.vector( dbar %*% solve(A1 + A2, dbar ) )
  a1inv <- Rfast::spdinv(A1)     ;     a2inv <- Rfast::spdinv(A2)
  mc <- solve( a1inv + a2inv, a1inv %*% ybar1 + a2inv %*% ybar2 )
  mc1 <-  - ybar1 + mc      ;      mc2 <-  - ybar2 + mc
  x1 <- Rfast::eachrow(y1, mc1, oper = "+")     ;     x2 <- Rfast::eachrow(y2, mc2, oper = "+" )
  B <- round( sqrt(R) )     ;     tb <- matrix(0, B, B)    ;    bm1 <- bm2 <- matrix(nrow = B, ncol = p)
  vb1 <- vector("list", B)    ;    vb2 <- vector("list", B)
  tb <- matrix(0, B, B)    ;    sqn1 <- sqrt(n1)      ;     sqn2 <- sqrt(n2)
  for (i in 1:B) {
    b1 <- sample(1:n1, n1, replace = TRUE)    ;    b2 <- sample(1:n2, n2, replace = TRUE)
    yb1 <- x1[b1, ]   ;   yb2 <- x2[b2, ]
    bm1[i, ] <- Rfast::colmeans(yb1)      ;       bm2[i, ] <- Rfast::colmeans(yb2)
    vb1[[ i ]] <- (crossprod(yb1) - tcrossprod( sqn1 * bm1[i, ]) ) / n1
    vb2[[ i ]] <- (crossprod(yb2) - tcrossprod( sqn2 * bm2[i, ]) ) / n2
  }
  for (i in 1:B) {
    for (j in 1:B) {
      vb <- vb1[[ i ]] + vb2[[ j ]]     ;    db <- bm1[i, ] - bm2[j, ]
      tb[i, j] <- db %*% solve(vb, db)
    }
  }
  ( sum(tb > test) + 1 ) / (B^2 + 1)
}
```